\documentclass[12pt]{article}
\usepackage{tabularx, booktabs, longtable, array, url, ltablex,graphicx}
\usepackage[normalem]{ulem}
\useunder{\uline}{\ul}{}
\usepackage{authblk}
\keepXColumns
\newcolumntype{L}[1]{>{\raggedright\arraybackslash}p{#1}}
\usepackage[colorlinks,allcolors=blue,urlcolor=blue]{hyperref}
\usepackage[acronym]{glossaries}
\glsdisablehyper 
\newacronym{llm}{LLM}{Large Language Model}
\newacronym{ai}{AI}{Artificial Intelligence}
\newacronym{hcai}{HCAI}{human-centered AI}
\newacronym{ir}{IR}{Information Retrieval}
\newacronym{nlp}{NLP}{Natural Language Processing}

\newcommand{\sect}[1]{Section~\ref{#1}}
\newcommand{\fig}[1]{Fig.~\ref{#1}}

\newcommand{\tab}[1]{Table~\ref{#1}}
\newcommand{\appx}[1]{Appendix \ref{#1}}

\usepackage[numbers,sort&compress]{natbib}
\bibpunct[, ]{[}{]}{,}{n}{,}{,}

\title{From keywords to semantics: Perceptions of large language models in data discovery}
\author[1,*]{Maura E Halstead}
\author[2]{Mark A. Green}
\author[1]{Caroline Jay}
\author[1]{Richard Kingston}
\author[1]{David Topping}
\author[2]{Alexander Singleton}
\affil[1]{University of Manchester, UK}
\affil[2]{University of Liverpool, UK}
\affil[*]{Corresponding author: maura.halstead@manchester.ac.uk}

\providecommand{\keywords}[1]{\textbf{Keywords: } #1}

\begin{document}
\maketitle 
\begin{abstract}
  Current approaches to data discovery match keywords between metadata and queries. This matching requires researchers to know the exact wording that other researchers previously used, creating a challenging process that could lead to missing relevant data. Large Language Models (LLMs) could enhance data discovery by removing this requirement and allowing researchers to ask questions with natural language. However, we do not currently know if researchers would accept LLMs for data discovery. Using a human-centered artificial intelligence (HCAI) focus, we ran focus groups (N = 27) to understand researchers' perspectives towards LLMs for data discovery. Our conceptual model shows that the potential benefits are not enough for researchers to use LLMs instead of current technology. Barriers prevent researchers from fully accepting LLMs, but features around transparency could overcome them. Using our model will allow developers to incorporate features that result in an increased acceptance of LLMs for data discovery.
\end{abstract}
\glsresetall

\glsresetall
\section{Introduction}
When research is conducted, scientists are encouraged by different stakeholders (e.g., policy makers, systems designers, librarians, funders, journals, other researchers, etc.) to openly share the resulting data so that it can be reused \citep{Mol2011plos-bio, GreGroSch2020Harvard}. When providing these data, researchers are expected to also supply metadata with carefully crafted \emph{keywords} in the titles, abstracts, and a few actual words. \emph{\gls{ir}} algorithms (e.g., Boolean, Vector Space \citep{SalWonYan1975acm-comm}, Term Frequency–Inverse Document Frequency \citep{Jon1972JoD}) depend on these keywords to search through the ever-growing number of datasets \citep{ChapSimKoe2020VLDB}. Recognizing the importance of metadata quality for the performance of \gls{ir} algorithms, stakeholders have called to make datasets more FAIR (``Findable,'' ``Accessible,'' ``Interoperable,'' and ``Reusable'') \citep{WilDumAal2016sci}. However, metadata quality does not appear to influence whether a dataset is used or not \citep{QuaRaf2022jis, Qua2023jis}. Therefore, what determines if a dataset is used or not could come from how users search for them (via locations or stringing together queries) or how the algorithm itself functions.

Users reuse datasets to support (e.g., evaluating and testing algorithms or methods, verifying results, teaching, etc.) and create new (e.g., modeling system inputs, testing new questions, replication, etc.) research \citep{GreCouGro2020jis, ChapSimKoe2020VLDB, KoeKacTen2017acm, GreGroSch2020Harvard}. To find this \emph{secondary data}, they go to multiple sources, such as, Google, data catalogs, government websites, internal sources, through research articles, and rely on social interactions in their own networks \citep{GreCouGro2020jis, KoeKacTen2017acm, PasBorWof2019harvard, KraPapCar2021ijdl}. Their strategy for finding datasets involves carefully selecting keywords they \emph{expect} researchers used in metadata. They will gain these keywords through using data and/or subject terms, browsing through websites, and examining references \citep{GreCouGro2020jis, KraPapCar2021ijdl}. Selecting these keywords is an important task, and if the \emph{right} keywords are not selected, the user could miss out on the most appropriate datasets. As a result, this search process is complex, challenging and time-consuming \citep{KoeKacTen2017acm, ChapSimKoe2020VLDB, KraPapCar2021ijdl,GreGroSch2020Harvard}. It also results in many datasets being \emph{invisible} and going underutilized \citep{QuaRaf2022jis, Qua2023jis}.

\glspl{llm} might be the technology that makes the search process easier and potentially eliminates the dependence of IR algorithms on metadata quality \citep{BerNevSko2022kis}. They are complex natural language processing models built using \gls{ai} methods and trained on massive amounts of datasets. Their training maps the relationships and meaning of terms within paragraphs and sentences, rather than merely matching keywords \citep{SunYanMa2024arXiv}. As a result, they allow users to write text-based queries and provide human-like responses \citep{LiTanZha2022arXiv, YanJinTan2023arXiv}. As such, \glspl{llm} may improve the quality of these search tools due to their ability to encode the nuances, context, and semantics of languages. This encoding means that they can handle complex and ambiguous queries, which is useful when users do not exactly know what keywords are most appropriate. The use of natural language queries may also be easier, thus simplifying the data discovery process.

While promising, the use of \glspl{llm} across society is not without controversy (e.g., hallucinated results, information bias, privacy and data risks, etc. \citep{ZhuFenXue2024jmed}). As such, the adoption of this technology is unlikely to be straightforward. It is currently unknown whether researchers would find these tools acceptable. Understanding their views and needs would help to produce tools that allow researchers to make the most of these emerging technologies for data discovery. As such, this research aims to understand researchers' potential use, perceptions, and requirements for using \glspl{llm} to search for and retrieve secondary data. The results of this study will benefit \gls{hcai} by providing a conceptual model that represents how researchers are likely to use \glspl{llm} for data discovery \citep{Shn2020ais-trans-hci}. They will also allow developers to incorporate the identified user requirements into \glspl{llm} so that researchers will use them for data discovery.

\section{Methods}
\subsection{Study design}
We conducted a qualitative study to examine the acceptability of \glspl{llm} for data discovery among users of secondary research data (pre-registration via OSF \url{https://doi.org/10.17605/OSF.IO/SB2MR}). We selected focus groups for data collection, as we wanted to capture diverse views. Ethical approval was acquired from University of Liverpool's Research Ethics committee (ID = 14290). Focus groups occurred online using Microsoft Teams. Each focus group was separated by user type, each lasted one hour, and all were conducted on 02 December 2024. Participants were reimbursed for their time. 

\subsection{Participants}
We identified four main types of research data users: (i) Doctoral students, (ii) researchers based at academic institutions (at any level), (iii) people involved in running data services, and (iv) researchers and analysts in local government, non-governmental organizations or third sector \citep{GreGroSch2020Harvard}. We sought to recruit up to 10 participants of each user-type across different disciplinary backgrounds. Participants were recruited by sharing adverts across the authors' networks, social media, and mailing lists. 

We recruited a total of $27$ participants, with seven doctoral students, five academic researchers, eight data service staff, and seven local government or third sector researchers. Of the participants in an academic setting, most studied Social Sciences ($N = 6$). Meanwhile, participants in industry roles mostly came from Education ($N = 5$) and Public Administration ($N = 5$) (see \tab{appx:tab:demo} in \appx{appx:part} for more details). 

The majority of participants indicated they had 2 to 5 years of experience using secondary data ($N = 9$), used it most of the time ($N = 9$), and were equally very and moderately familiar with secondary data ($N = 7$). No participants indicated that they were not familiar with secondary data. Most participants indicated that they used secondary data to explore new research questions ($N = 17$). Industry professionals wrote that they used secondary data to support researchers and gain a better understanding of what they needed, assess the feasibility of research proposals, and ensure that access to the data was ``safe'' (see \tab{appx:tab:demo} in \appx{appx:part} for details).

\subsection{Materials}
\begin{table}[ht]
  \centering
\caption{Focus group questions}
\label{tab:fgQuestions}
{\scriptsize
  \begin{tabular}{p{0.4\textwidth}  p{0.4\textwidth}}
\toprule
\textbf{Main Question} & \textbf{Follow-up} \\
\midrule
How do you currently search for existing data? & Scope, search strategy, or refining and assessing the results. \\

How would you search for existing data with an LLM? & What do you think are the differences between how you would use and what you found from using keywords or questions? \\

For the LLM, what other information (besides the dataset itself) do you hope to obtain when you search for existing data? &  \\

What do you think are the strengths, weaknesses, opportunities, and threats of using LLMs to search for secondary data? & Would you find it reliable for finding datasets for secondary research? What information would you like to obtain to know the results are valid? \\

When it comes to dealing with secondary data, what would you most like the LLM to have or be able to do? &  \\
\bottomrule
\end{tabular}
}
\end{table}

\tab{tab:fgQuestions} displays the main and follow-up questions asked of the participants. As an introduction, participants were given a scenario to prompt their thoughts about searching for secondary data. This scenario was that they wanted to find secondary data around fuel poverty and associated issues. The first two questions were the most directed questions as we wanted to understand how secondary data discovery might be related to or differ based on which type of query method was used. The remaining questions were more open-ended. To answer the third question, participants used a Canva whiteboard to write down their feelings towards \glspl{llm}. The last question was to close the session and summarize the important characteristics for an \gls{llm} to have if used for data discovery.

Participants also answered demographic questions via Qualtrics (see \appx{appx:demo:survey} for details). 

\subsection{Data Analysis}

We followed a six-step inductive thematic analysis \citep{NaeOzuHow2023ijqm}: familiarity with transcript, turn quotes into keywords, group keywords into codes, group codes into sub-themes and themes, conceptualization, and development of a conceptual model. Two members of the team followed step 1 and became familiar with the transcript. One member of the team followed steps 2--4 to perform the thematic analysis two times (the first was done in February 2025, the second in May 2025). These were then reviewed by the second member of the team and a final set of themes were agreed. All members of the team, coming from diverse disciplines, checked the themes and finalized the conceptualization. 

\section{Results}

\begin{figure}
\centering%
\includegraphics[width=0.5\textwidth]{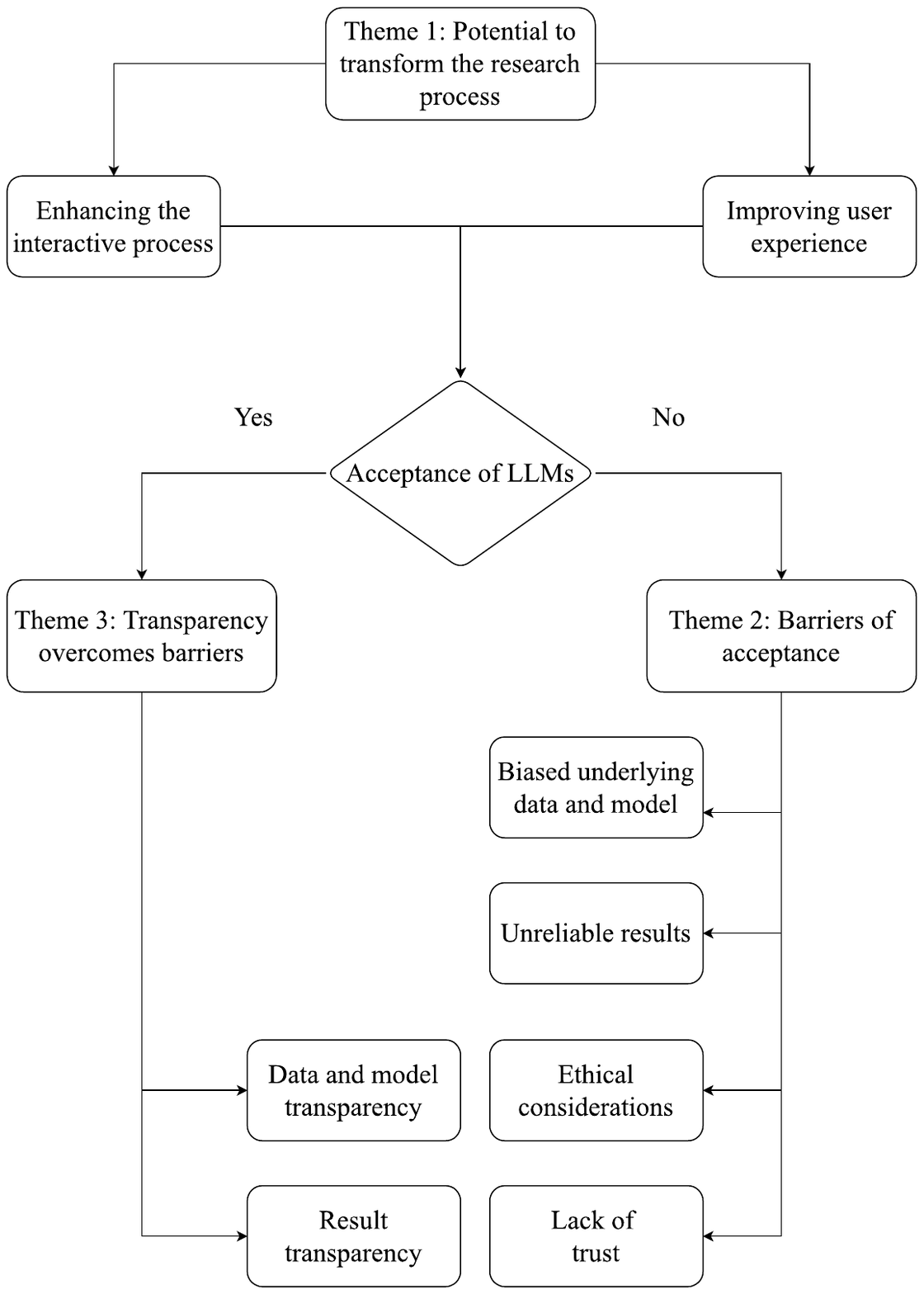}
  \caption{Conceptual Model of Themes} 
  \label{fig:results}
\end{figure}

Coded files are shared openly via OSF \url{https://osf.io/fntyd}. \fig{fig:results} displays the conceptual model representing the use of \glspl{llm} for data discovery. As shown, three themes emerged: potential to transform the research process, barriers to acceptance, and transparency overcomes barriers. Participants thought that \glspl{llm} could streamline the process and experience of data discovery. While the promise of \glspl{llm} could make research easier, barriers could prevent researchers from fully embracing them. Transparency can overcome these barriers and lead researchers to adopt \glspl{llm} for data discovery. 

\subsection{Theme 1: Potential to transform the research process}\label{res:themes:1}
During the research life cycle, finding the right datasets is an iterative process in which researchers formulate a research question, choose where to search, construct a search query, filter the results, and ultimately select a dataset. This process closely resembles the systematic approach for performing literature reviews \citep{BraDeJRet2018jmla, KraPapCar2021ijdl}. In this theme, we explore how participants currently use keyword-based search and the potential for \glspl{llm} to transform both the procedure and the researcher’s experience. 

\subsubsection{Enhancing the interactive process}\label{res:themes:1:process}
\begin{table}
  \centering
  \caption{Theme 1, Interactive process topics.}
  \label{tab:theme1:interactive}
  {\scriptsize
  \begin{tabular}{p{0.1\textwidth} p{0.1\textwidth} p{0.3\textwidth} p{0.4\textwidth}}
\toprule
\textbf{Topic} & \textbf{Query Type} & \textbf{Findings summary} & \textbf{Example quotes} \\
\midrule

Research question formation & Keyword & Start with solidifying research question before searching & ``Just sort of matching the information we have to the information we need and seeing if there are any data gaps there.'' \newline
``I would start by trying to put together the research question in my head before I started Googling on this.'' \\

& LLM & Helps identify research question before searching
& ``Could help researchers to compose a research question.'' \newline
``Identify gaps in research - so you know your research is unique.' \\

Query strategy &  Keyword
& Selectively match a few keywords to research question with a trial and error approach. &
``I would probably start with putting some keywords around the question.'' \newline
``I would just provide the key words and no filler words.'' \\
& LLM & Can use natural language in a conversation-like style. Start with the end in-mind.
& ``I would phrase it in terms of the outcomes I was trying to achieve, almost typing the research question instead of thinking this is the data I need. It is start with the end in mind. I am looking for data to solve a question, and then let the system suggest the datasets back to me.'' 
\\

Search locations & Keyword & Everyone uses Google. Research question determines which data catalogue to use. Use networks to identify data. & ``Yeah, I would just initially go onto my search engine and do a Google search… But almost always I am probably going to start there if I am perfectly honest.'' \\

& LLM & Depends on desired outcome.
& ``...So it really depends on time and who I am reporting this to and what the end results are that I need.''
\\

Filtering findings & Both & Review variable level information, prioritize document source (e.g., journal article, report, blog, etc.), which requires metadata. Also examine citation metrics. & ``Yes, I think that step I would be trying to understand what the datasets contained. So as everyone said, reading the metadata, I know the description or abstract can be quite useful, but sometimes very difficult to understand exactly what is contained within the dataset, so trying to get a handle on what information the data actually contained to see if it is relevant to answering the question.'' \newline
``If there is a source that has been highly cited, especially if it has been published recently, that would suggest to me that it is worth looking at, if not, reliable.''\\

What is found & LLM & Provide new insights, inspire new research, get quick summaries. Mixed views about data popularity: find under-utilized data or only popular data. &  ``Provide other ideas that can be developed in qualitative research, potentially even if I am looking for other data to complement it.'' \newline
``Identifying data that otherwise might not have been used.'' \\
\bottomrule
  \end{tabular}
  }
\end{table}

\tab{tab:theme1:interactive} displays the topics within this sub-theme, with a summary of the findings and some example quotes from the participants. When using keywords, participants explained that they need to first narrow down their research question to pick a few prominent ones for their query. They explained that this process is challenging and they would have to use `trial-and-error' until they started seeing what they were intending to locate, repeated across different sources. In comparison, participants thought that using an \gls{llm} could help them identify their research question by using natural language to search for data. 

The location of where participants went to search for data with keywords depended on their research question. However, almost all participants agreed or mentioned using Google to locate data. People in all focus groups discussed the importance of asking colleagues, subject-area experts, and their existing networks when they were not sure which data to use. Human contact was viewed as a way of identifying trusted sources. 

When filtering through all the datasets found from a query, participants often found it difficult or even overwhelming to understand the amount of information available for reviewing the relevance of datasets. They explained that variable-level information that includes completeness helps narrow down which datasets to inspect further. They also prioritized the document source (e.g., journal article, report, blog, etc.) and like to review the citation metrics to get an idea of how the dataset has been used elsewhere. They thought that \glspl{llm} could streamline this process by summarizing complex documentation. Participants also thought that \glspl{llm} could tell the user how the dataset relates to their query and provide statistics for the variables in the dataset. Participants thought that it would be useful for the \gls{llm} to report the supporting documentations' completeness, list the pros and cons of the dataset, and suggestions on how to use a dataset (e.g., requirements to have specific software). 

There was not much discussion about what participants got out of using keyword searches. Instead, they focused on what \glspl{llm} could offer in terms of what they found. Participants thought that \glspl{llm} can help to create new insights and inspire new research ideas. They also thought that \glspl{llm} might aid in finding other relevant topics. In terms of locating specific datasets, some thought that \glspl{llm} have the potential to find under-utilized data sources and identify data that might not have been used otherwise. 

\subsubsection{Improving user experience}\label{res:themes:1:experience}
\begin{table}
  \centering
  \caption{Theme 1, Improving user experience topics.}
  \label{tab:theme1:userExperience}
  {\scriptsize
  \begin{tabular}{p{0.2\textwidth}  p{0.3\textwidth} p{0.4\textwidth}}
\toprule
\textbf{Topic} & \textbf{Findings summary} & \textbf{Example quotes} \\
\midrule

Customization & Provides flexibility in terms of what the LLM searches through & ``Can limit training of model to relevant internet sources e.g. legit data catalogs and repositories '' \\

Efficiency & Speeds up the process of data discovery. & ``I guess the benefit of LLM search tools is that they combine many databases of information so you are not having to search multiple times, keywords, multiple times across sort of six different databases. That you can just do it that once. '' \newline 
``Particularly in the context of working in a trusted research environment, where sort of both the external researcher when they are developing and submitting their project application, they will be required to search through metadata, try to identify the right datasets, the right variables for their project, and then likewise on our side as well. Internally, we will have to review those project applications and do the same thing. Have a look at which datasets, which variables they have specified, which can be quite time consuming. Not always easy if the sort of metadata is not saved correctly, things like that. So something like this just on both sides, I think has a potential to speed up that process.'' \\

Users & Makes data more easily accessible to those not in expert domains or native English speakers. & ``Gain an understanding of the subject matter requirements without being a subject matter expert.'' \\

Service & Want a user-friendly experience, guidance on how to use it efficiently, and how to judge what the LLM finds. & ``I am thinking it as a service basically. So what I would want for it is personally for most be user friendly so I could use it so I would at the end of the day it's going to give me suggestions. So I want to be using it easily and they have the relevant information that we discussed before so it could help me make an informed decision.'' \\
\bottomrule
  \end{tabular}
  }
\end{table}

\tab{tab:theme1:userExperience} shows the topics coded into this sub-theme with summaries and example quotes. When going through the research process, the user needs to be able to efficiently navigate resources. Almost all of the participants mentioned that the main benefit of using \glspl{llm} in data discovery was that LLMs would speed up the process of finding relevant data. LLMs can quickly analyze data quality and assess whether each dataset would answer the research question. Furthermore, \glspl{llm} can divulge any data permission issues before the user tries to analyze the data. Participants still want to have a search results page so that they can decide which dataset to use themselves, instead of the \gls{llm} making that decision for them.

Participants also thought that \glspl{llm} can be customized to only look at specific catalogs, allowing for a more personalized search experience. In terms of users, participants stated that \glspl{llm} can increase inclusivity by allowing non-native English speakers and non-specialists/experts to access data. These users, along with the `typical' researcher, need to have a user-friendly experience. As such, there needs to be training and the user should be empowered to validate what the \gls{llm} finds.

\subsection{Theme 2: Barriers to acceptance}\label{res:themes:2}

\begin{table}
  \centering
  \caption{Theme 2, Barriers to acceptance.}
  \label{tab:theme2}
  {\scriptsize
  \begin{tabular}{p{0.1\textwidth}   p{0.2\textwidth} p{0.6\textwidth}}
\toprule
\textbf{Sub-theme} & \textbf{Findings summary} & \textbf{Example quotes} \\
\midrule

Biases in underlying data and overall &  Data quality determines the quality of outputs. Lack of access to predigitized information. Popular datasets might be always returned. Bias might be compounded based on reinforcing data choices. The way the model works makes it difficult to reproduce what LLMs find.

& ``The underlying structure of the LLMs is influenced largely by what is available online and that captures certain types of data, but not everything. So pre-digitised information that is stored in physical form, in a library or indigenous knowledge, or oral culture and traditions, are not well represented. You are not starting with a sum of the entirety of human knowledge. It is very much English language-based information from 1980 onwards.” \newline ``I was thinking of the same thing that the language models are training themselves and they could train themselves at the wrong way to focus in on rather than find the unusual, which is what we are really probably after.'' \newline
``I was just thinking about this that if the large language models are trained on particular types of datasets and they have particular inputs, you might get an output which is true. Like it is accurate, it is pointing to real things, but it is pointing you to things that are built on the structures that it sees underneath. So I mean, we have talked about citations, we have talked about where data is found. We might think that those are places we want to put our trust in, but if those are skewed by other biases, so for instance, citation metrics have been kind of data mined, you know, to manipulated, a human being may evaluate that in one way, but the machine may evaluate it in another. And if we use the machine, then we are kind of going in a particular direction. If we are all going in that direction, then we have compounded an underlying bias in the system.'' \newline
``That you can rephrase the same query in very, very marginally different ways and get very different responses. Sometimes dramatically opposite responses...'' \\

Unreliable results & The LLM's tone is convincing, even when wrong. LLM might create fake data. &  ``They [LLMs] are always happy, always positive, always agreeing with you. It basically could lead you like into an echo chamber and could mislead you towards a place that is not the place that you want to be. Only because it does not challenge you. So a part of critical thinking is to be able to be challenged and think out of it.'' \newline ``...Not only was the data misrepresented, but sometimes they would be made-up in that sense to explain what the dataset is and exactly what it provided.'' \\

Ethics & If LLMs provide different quality, those from developing countries could be left behind. Users could provide sensitive information into queries and, if hacked, this information could be leaked. LLMs can take ideas from sources without providing credit. Unclear environmental costs causes users to avoid LLMs. &
``... when you use nonpaying versions that they come with very low quality of information. So I think that is a risk actually for our developing countries, it is not easy to use these tools and they get something with the quality information.'' \newline
 ``Depending how the model is implemented, some of them will be harvesting your search intents and any data or prompts you put in it so. You have to be careful, you might be exposing sensitive information or confidential information.'' \newline
 ``When you get those outputs that are kind of a whole sentence or a whole structure, a whole paragraph, it is building on data and it is building on information, but it does not have the scholarship necessarily to be able to say this comes from here. And so you either have to disentangle that yourself or you are at risk of using other people's ideas without being able to give appropriate recognition to who did it and so effectively, you are accidentally plagiarizing people all the time.'' \newline
 ``And then I would probably go on Google for any gaps trying to limit how much is done in the LLM because of, just thinking about environmentally conscious... how much more energy and waters consumed by using an LLM versus using a Google query.''\\
 Lack of trust & LLMs used to validate other findings. Cannot trust what LLMs find and must verify with other search engines.

& ``...what I would probably do is just ask it the same question multiple times, perhaps wording it slightly differently each time and then see what comes up repeatedly. But I would still double check it and verify anything that came back before I trusted it.'' \\
\bottomrule
  \end{tabular}
  }
\end{table}

\tab{tab:theme2} shows the finding summaries for the sub-themes in Theme 2 along with example quotes. While \glspl{llm} have the potential to revolutionize the data discovery process and experience, there are some fundamental concerns that prevent users from fully embracing them. Participants raised concerns about bias that could mislead the user and how the results found can be unreliable. They also voiced some ethical considerations and a lack of trust. 

\subsubsection{Biases in the underlying data and overall}
Data used to train the \gls{llm} helps to determine what it finds. This process makes the quality of the underlying data vital. As such, participants thought that poor quality data placed into an \gls{llm} would cause it to come back with bad information. They discussed that a part of this poor quality data comes from incomplete, missing, or inefficiently parsed datasets. 

Contributing to incomplete data are censored datasets. These types of datasets are ones where some data is omitted. Therefore, it only shows part of a dataset, not all of it. So, if the \gls{llm} only has access to part of a dataset, it does not know the ``whole truth.'' As for missing data, the \gls{llm} most likely does not have all data that has ever existed. Furthermore, older digitized datasets are less standardized than modern ones, causing the LLM to misinterpret what they contain. 

Despite some participants acknowledging that \glspl{llm} can find under-utilized datasets (see \ref{res:themes:1:process}), others thought that popular datasets might be prioritized by the model. As researchers are interested in finding knowledge gaps, this approach could hinder research instead of help it. Furthermore, favoritism of popular datasets could lead to compounded biases that influence what future searches return (e.g., the same datasets found instead of those that are less used and could better answer the research questions). Thus, the researcher's choices could influence the \gls{llm}, and vice versa, causing biases to be reinforced within a feedback loop. 

\subsubsection{Unreliable results}
Biases from the underlying data or model could mislead the user into drawing incorrect conclusions. For example, if the \gls{llm} does not have all the data or focuses on a few datasets instead of everything, it cannot find the most appropriate datasets for a query. Participants explained that a reason users believed what the \gls{llm} finds is due to its convincing and authoritative tone. Since the \gls{llm} does not challenge the user or explicitly state that it is unsure, the information seems believable, even when it is wrong.

These types of responses are especially problematic if the \gls{llm} hallucinates or generates fake datasets. Participants felt that these hallucinations can sound convincing and trick the user into believing that they are real. 

\subsubsection{Ethical considerations}\label{res:themes:2:ethics}
As noted in \sect{res:themes:1:experience}, an agreed-upon benefit of using \glspl{llm} for data discovery was that diverse user types could now search for data. However, one participant warned that having free and paid versions would disproportionately affect those from developing countries that cannot afford to pay for higher quality answers. Consequently, those users could be left behind, with lower quality information about research outputs. 

In terms of user actions, participants noted the potential for ``hackers'' that could try to access sensitive information. If a user types sensitive information into a query, it could be unintentionally harvested from a motivated intruder. Using their current understanding of LLMs and not necessarily how they might be used for data discovery, participants thought that using an \gls{llm} could result in unintentional plagiarism. This plagiarism arises because there is currently uncertainty around who the original authors are in terms of ideas and voice. LLMs should make it clear who the credit belongs to, when finding datasets.

Environmental concerns were also mentioned as there is not enough transparency about what it costs \glspl{llm} to run queries. Sustainability was an important topic to many participants, and they want to be able to decide whether the utility of making searches easier with \glspl{llm} outweighs the environmental cost.

\subsubsection{Lack of trust}
Participants indicated that they did not fully trust \glspl{llm}. This lack of trust was evidenced through their discussion of using multiple \glspl{llm} with multiple versions of their queries. It was also exhibited through their requirement to ``double-check'' anything found using \glspl{llm}. Inconsistent responses also do not help participants trust \glspl{llm}. This lack of trust makes participants only use \glspl{llm} to validate what they found with other tools. 
 
\subsection{Theme 3: Transparency overcomes barriers}\label{res:themes:3}

\begin{table}
  \centering
  \caption{Theme 3, Transparency overcomes barriers.}
  \label{tab:theme3}
  {\scriptsize
  \begin{tabular}{p{0.1\textwidth}   p{0.2\textwidth} p{0.6\textwidth}}
\toprule
\textbf{Sub-theme} & \textbf{Findings summary} & \textbf{Example quotes} \\
\midrule
Data and model transparency & Want to know that the information about training data, details of the model, and explicit limitations. Having these details will help in reporting searches to others. Transparency could also make researchers produce better documentation if they get feedback from the LLM that their dataset is not being returned due to documentation weaknesses. Knowing the model's confidence in its finding, and how the model parses the query and finds datasets could also help increase trust. &  ``Yeah, I think these kinds of systems should be designed to be responsible. And what I mean by that, is that they should all fit all the sort of caveats that we have identified and been talking about. I think anytime you get search results, it should be quite clear about what the limitations are of what it is offering you. And I think presenting that information alongside search results would go a long way towards helping people use these kinds of systems confidently...'' \newline
``Yes, I do not know if this is possible, but in fact could be some sort of confidence score associated with the response. Because, yeah, the experience sort of large language models telling me very confidently that, you know, I have asked the question, here is the answer. And then I sort of go off and interrogate that answer and find out it is completely made-up. So if there is some way of saying, well, here is what I feel the LLM saying here is the answer. But I have got low confidence in it, or I have got very high confidence in it. And if I can be assured that, yeah, that is really the case and something like that would be great.''
\newline ``Transparency about how it reached the the answer. Essentially something like this. You know, we looked at, an X number of datasets and I think of these are the top five I would suggest, but if you want to go further do this. So that kind of like descriptive analysis around how the search has been done would be good.'' \\

Response transparency & Provides easy access to download the dataset and look at the metadata. Allows researchers to explore where else the data has been used and what for. & ``I think direct access to downloading the datasets the that you select, because simply telling you that the datasets are there still leaves a lot of grunt work before you can set up your own analytical framework. If there was some more standard way of being able to download the data, load the data into your own little analytical framework that will be helpful.'' \newline
``I think just as a researcher, I would love to know what is out there and who has used it before, just so that I can get a sense of what the academic literature has prioritised. And even if you do not choose that, it will be good to know why you do not choose the second data. So I think just having the data there and then knowing who has used it would be really useful. ''
\\
\bottomrule
  \end{tabular}
  }
\end{table}

\tab{res:themes:3} displays the sub-themes and summaries with participant quotes. A way to overcome the barriers mentioned in \sect{res:themes:2} is to provide transparency. Participants felt that transparency is required in terms of the data, model, and responses. Without this transparency, they will not accept technologies that completely replace keyword search with \glspl{llm}.

\subsubsection{Data and model transparency}
Participants require transparency about where the \gls{llm} looks for data and details around the model itself. They wanted to know that the data were coming from reliable sources. When discussing the model, participants wanted clear and concise limitations. Participants thought that being upfront about limitations was responsible and would make the system more trustworthy. They also wanted to know details like model version number, release dates, and who is responsible for its development and oversight. This type of information makes it easier for researchers to report their findings to other researchers.  

Another area discussed around model transparency was the \gls{llm} reporting its own confidence about answering queries. Participants felt that the system should be able to articulate if it or the data was uncertain. They also wanted more explainability around how the \gls{llm} used the query to find datasets.

\subsubsection{Response transparency}
Having a direct link to the data and metadata was one of the most common requests from all user groups. Participants felt this was the easiest way to know that the data was valid and not hallucinated. As discussed in \sect{res:themes:1:process}, when using keyword search, an important part of selecting which datasets to use requires examination of metadata. When examining the metadata, participants also investigate the citation metrics to see where else the data was used. Participants still find this process helpful and want \glspl{llm} to have similar features. They still want to explore the citation metrics when using \glspl{llm}. \glspl{llm} could also be used to give an overall summary of this information initially, to help researchers on where they should focus their reading of this information. 

\section{Discussion}
This study used focus groups to explore the suitability of \glspl{llm} for data discovery with research data users. Our conceptual model, informed from our themes, shows that the benefits of using \glspl{llm} are not the only determinate of whether researchers will use them for data discovery. Although LLMs bring much-awaited features into the data discovery process, there are several barriers and concerns that can limit their use or acceptability. Majority of these barriers can be overcome with transparency. Our findings help to identify actionable strategies that can help to reassure researchers and set out best practice for designing LLM-augmented tools.  

\subsection{Contributions and Implications}
Our findings offer a range of contributions to the field of \gls{hcai} \citep{Shn2020ais-trans-hci}. With different participants and a research question, our findings on how researchers use keyword search closely follow previous research \citep{KoeKacTen2017acm, ChapSimKoe2020VLDB, KraPapCar2021ijdl,GreGroSch2020Harvard}. Since these findings are repeated, it shows the appropriateness of our sample and that our understanding of the keyword search process is reliable. However, we found that participants would use LLMs differently. This finding indicates that our understanding of the data discovery process is likely to change with the incorporation of this technology. As such, the user requirements will also evolve. So, if we want researchers to abandon what they are currently comfortable with for a new procedure that provides better value over the current one \citep{GreHinStr2010bmj, KimKan2009MISQ}, we need to understand their requirements for using LLMs \citep{BinCurLoc2023comphb, Shn2020ais-trans-hci}. 

Participants identified a range of values that LLMs may bring to enhance the data discovery process. These included facilitating more natural queries, supporting non-native English speakers or non-specialists who may not know the right terminology to use. Participants thought that the most powerful part of LLMs could be used to provide summaries of datasets to help them decide which dataset to use, an important feature currently missing \citep{KoeSimBlo2020ijhcs}. They also thought that summaries could break down complex metadata and documentation that is often challenging to understand \citep{KraPapCar2021ijdl}. The incorporation of additional services offered by LLMs along with finding relevant datasets could help to re-conceptualize the interface between researchers and data infrastructure, creating tools that are more dialogic, context-aware and tailored to individual needs.  

Despite these benefits, participants indicated hesitation to accept LLMs due to underlying biases, technical issues like tone and hallucinations, and ethical concerns, following previous discussions around the general use of LLMs \citep{NahZheCai2023JITCAR}. Focusing on data discovery, our participants thought that there might be biases in the underlying data and LLM itself. However, these same participants unashamedly promoted the use of Google even though it is also known to have these types of biases \citep{HabStoHig2024ARXIV}. Another concern was that the tone offered by LLMs could cause users to trust the findings even when incorrect. This concern is valid as persuasive text does lead to an increase in sharing misinformation \citep{ZhoLiLu2021IPM}. Participants also expressed ethical concerns like alienating poorer countries, security breaches, and unclear environmental costs. 

While these barriers appear large, participants described how they could easily be overcome by providing transparency \citep{AtfLew2025ieee-trans-techsoc}. Building trust requires tailoring systems to the users specific needs \citep{BacKhaHal2022ijhci}. Our users discussed wanting information regarding training data, model, and uncertainty. They also want to have direct links to datasets and explicit limitations regarding the model and data. 

In line with the idea that \gls{ai} should not replace but instead support users \citep{Shn2020ais-trans-hci}, there was little interest in the automation of the data discovery process. Human judgment was still critical for selecting datasets. This suggests that tools which complement, rather than replace, human expertise and agency are most valuable to researchers. This view has been demonstrated across other technology and points towards the importance of \gls{hcai} \citep{Shn2020ijhci}. For example, participants wanted LLMs to identify relevant data, but not to make final selections on their behalf \citep{LubTan2019advnips}. 

Our findings suggest that researchers do not expect perfect AI or LLM tools. They understand both the value that they may bring to the research process, as well as their limitations. Rather, researchers want transparent, accountable and explainable tools that can augment their current approach to the data discovery process or extend current services. 

\subsection{Limitations and Future Research}
With our focus group design, we have a few limitations. First, we targeted focus groups based on those from Gregory et al. \cite{GreGroSch2020Harvard}. However, in our analysis, we noticed that participants identified as both government or third sector and data service staff. In hindsight, this joint identity makes sense. For example, people who work for ``Office of National Statistics'' are part of a government organization, but they can also work as data support for their catalog system. Thus, we have unbalanced groups where these two research roles are overrepresented. Second, at the time of data collection, there were no LLMs that specifically helped researchers find datasets. This novelty means that our participants had to imagine how they might be used in data discovery. Their imagination was most likely based on their current knowledge of using general LLMs like ``ChatGPT,'' ``Gemini,'' ``Copilot,'' etc. Therefore, future research should conduct similar focus groups with users that have experience using LLMs directly for data discovery. These focus groups could follow an agile approach to systematically discover user requirements that build trust \citep{HusSlaHol2009lecture, HolSarAng2022sensors}. They also would allow for a better direct comparison of using keyword and LLMs for data discovery.

\subsection{Conclusions}
In conclusion, whilst LLMs hold transformative potential for revolutionizing data discovery, there are challenges for ensuring that any tools are acceptable to use beyond just technical requirements. Our study demonstrates the importance of co-designing any LLM-augmented search tools alongside users. Where tools can address the limitations or openly report on the concerns outlined here, LLMs may help to not only improve data discovery but also reshape how researchers engage and interact with datasets towards a more inclusive, intuitive, and efficient approach to research.

\section*{Acknowledgements}
We want to thank all of the focus group participants for their time and discussions. We also thank the members of the projects external advisory board for their advice in designing the focus group and reviewing the findings. 
\section*{Disclosure of interest}
No potential competing interest was reported by the author(s).

\section*{Declaration of Funding}
This work was supported by the Economic and Social Research Council [grant number ES/Z502947/1]. 

\bibliographystyle{plainnat}  
\providecommand{\doi}[1]{}
\renewcommand{\doi}[1]{\hspace{.16667em plus .08333em}\discretionary{}{}{}\href{https://doi.org/#1}{\urlstyle{rm}\nolinkurl{#1}}}

\appendix
\section{Participants}\label{appx:part}
{\scriptsize
\begin{longtable}{>{\raggedright\arraybackslash}p{0.2\textwidth} 
                  >{\centering\arraybackslash}p{0.14\textwidth} 
                  >{\centering\arraybackslash}p{0.14\textwidth} 
                  >{\centering\arraybackslash}p{0.14\textwidth} 
                  >{\centering\arraybackslash}p{0.14\textwidth} 
                  >{\centering\arraybackslash}p{0.14\textwidth}}
\caption{Participant demographics.} \label{appx:tab:demo} \\
\toprule
\textbf{Demographic} & \textbf{PhD students} & \textbf{Academic Researchers} & \textbf{Government or third sector} & \textbf{Data service staff} & \textbf{Total} \\
\midrule
\endfirsthead

\toprule
\textbf{Demographic} & \textbf{PhD students} & \textbf{Academic Researchers} & \textbf{Government or third sector} & \textbf{Data service staff} & \textbf{Total} \\
\midrule
\endhead

\midrule
\multicolumn{6}{r}{\emph{Table continues on the next page}} \\
\midrule
\endfoot

\bottomrule
\endlastfoot

\multicolumn{6}{l}{\textit{Field of study}} \\
Social Sciences & 4 & 2 & NA & NA & 6\\
Health and Medicine & 1 & 2 & NA &  NA & 3\\
Business & 1 & 0 & NA & NA & 1\\
Engineering and Technology & 1 & 1 & NA & NA & 2 \\
& & & & & \\
\multicolumn{6}{l}{\textit{Industry}} \\
Public Administration and Governance & NA & NA & 4 & 1 & 5 \\
Healthcare and Social Services & NA & NA & 1 & 1 & 2 \\
Urban Planning and Development & NA & NA & 1 & 0 & 1\\
Education and Workforce Development & NA & NA & 0 & 5& 5\\
Environment and Sustainability & NA & NA & 1 & 1 & 2\\
& & & & & \\
\multicolumn{6}{l}{\textit{Years of experience}} \\
0--1 year & 0 & 0 & 1 & 4 & 5 \\
2--5 years & 6 & 1 & 2 & 0 & 9\\
6--9 years & 1 & 0 & 3 & 2 & 6\\
10+ years & 0 & 4 & 1 & 2 &  7\\
& & & & & \\
\multicolumn{6}{l}{\textit{Frequency}} \\
Never & 0& 0& 0& 3& 3\\
Sometimes & 0& 2& 1& 3& 6\\
About half the time & 2 & 1 & 1 & 0& 4\\
Most of the time & 4& 1& 4& 0& 9\\
Always & 1 & 1 & 1 & 2& 5\\
& & & & & \\
\multicolumn{6}{l}{\textit{Familiarity}} \\
Extremely familiar & 2& 1& 2& 1& 6\\
Very familiar & 3& 1& 1& 2& 7\\
Moderately familiar & 2& 2& 2& 1& 7\\
Slightly familiar & 0& 1& 1& 4& 6\\
Not familiar at all & 0& 0& 0& 0& 0\\
& & & & & \\
\multicolumn{6}{l}{\textit{Goals}} \\
Replicate or validate findings & 2 & 2 & 2 & 2 & 8  \\
Explore new research questions & 7 & 5 & 3 & 2 & 17 \\
Conduct preliminary analysis before primary data collection &  1 & 4 & 3 & 2 & 10 \\
Perform meta-analysis or systematic review &  2 & 2 & 3 & 1 &  8 \\
Other: &  0& 0 & 2 & 3 & 5 \\

\end{longtable}}

\section{Demographic Survey}\label{appx:demo:survey}
\begin{enumerate}
    \item What is your primary field of study? (only to PhD and academic researchers)
    \begin{enumerate}
        \item Social Sciences (e.g., Sociology, Psychology, Political Science, Economics, Geography, Education)
        \item Health and Medicine (e.g., Public Health, Epidemiology, Biomedical Research)
        \item Business (e.g., Management, Marketing, Finance, Accounting, International Business)
        \item Engineering and Technology (e.g., Computer Science, Data Science)
        \item Other (please specify):
    \end{enumerate}
    \item What is your primary industry? (only to government or third sector or data service staff)
    \begin{enumerate}
        \item Public Administration and Governance (e.g., Municipal Government, State or Provincial Government, Federal Government, Public Policy)
        \item Healthcare and Social Services (e.g, Public Health, Social Welfare, Healthcare Administration, Community Development)
        \item Urban Planning and Development (e.g, Urban Planning, Infrastructure Development, Housing and Real Estate)
        \item Education and Workforce Development (e.g., K-12 Education, Higher Education, Workforce Development, Vocational Training)
        \item Environment and Sustainability (e.g., Environmental Protection, Climate Policy, Sustainability Initiatives, Natural Resources Management)
        \item Other (please specify):
    \end{enumerate}
    \item How many years of experience do you have in research or data analysis?
    \begin{enumerate}
        \item 0 - 1 year
        \item 2 - 5 years
        \item 6 - 9 years
        \item 10 + years
    \end{enumerate}
    \item How frequently do you use secondary data in your work?
    \begin{enumerate}
        \item Never
        \item Sometimes
        \item About half the time
        \item Most of the time
        \item Always
    \end{enumerate}
    \item How would you describe your familiarity with secondary data?
    \begin{enumerate}
        \item Extremely familiar
        \item Very familiar
        \item Moderately familiar
        \item Slightly familiar
        \item Not familiar at all
    \end{enumerate}
    \item What are the main goals you aim to achieve with secondary data? (Select all that apply)
    \begin{itemize}
        \item Replicate or validate findings
        \item Explore new research questions
        \item Conduct preliminary analysis before primary data collection
        \item Perform meta-analysis or systematic review
        \item Other:
    \end{itemize}
\end{enumerate} 

\section*{About the authors}
\textbf{Maura E Halstead} is a Postdoctoral Research Associate at the University of Manchester. Her research focuses on human-computer and human-algorithm interaction, with a specific interest in decision-making and experimental design.
\newline
\noindent \textbf{Mark Green} is Professor of Health Geography at the University of Manchester. Their research focuses on how new forms of data can supplement traditional research datasets for studying the drivers of health inequalities. 
\newline
\noindent \textbf{Caroline Jay}  is a Professor of Computer Science and Head of Research in the School of Engineering 
at the University of Manchester. Her research focuses on software engineering and human-computer interaction, combining her expertise in computer science and psychology. She uses empirical 
methods to study and innovate in various areas.
\newline
\noindent \textbf{Richard Kingston} is Professor of Urban Planning and GISc, Director of the NERC Digital Solutions Programme, Director of UoM's Digital Futures Platform and Deputy Director of the Spatial Policy Analysis Lab. He joined the University in September 2003 and was previously a research associate at the Centre for Computational Geography, School of Geography, University of Leeds where he developed PPGIS methods and applications.
\newline
\noindent \textbf{David Topping} is a Professor of The Digital Environment at the University of Manchester, alumni of the Alan Turing National Institute for AI and a member of the Natural Environment Research Council Science Committee.
\newline
\noindent \textbf{Alexander Singleton}  is a Professor of Geographic Information Science at the University of Liverpool, where he was appointed as a Lecturer in 2010. Previously he held research positions at University College London, where he was also awarded a PhD in 2007. He completed a BSc in Geography at the University of Manchester, graduating with a First-class honours degree in 2003.
\end{document}